# An Efficient Authentication Protocol for Smart Grid Communication Based on On-Chip-Error-Correcting Physical Unclonable Function

Masoud Kaveh[1], Mohammad Reza Mosavi[2], Diego Martín[3]*, and Saeed Aghapour[4]

*Abstract*— Security has become a main concern for the smart grid to move from research and development to industry. The concept of security has usually referred to resistance to threats by an active or passive attacker. However, since smart meters (SMs) are often placed in unprotected areas, physical security has become one of the important security goals in the smart grid. Physical unclonable functions (PUFs) have been largely utilized for ensuring physical security in recent years, though their reliability has remained a major problem to be practically used in cryptographic applications. Although fuzzy extractors have been considered as a solution to solve the reliability problem of PUFs, they put a considerable computational cost to the resource-constrained SMs. To that end, we first propose an on-chip-error-correcting (OCEC) PUF that efficiently generates stable digits for the authentication process. Afterward, we introduce a lightweight authentication protocol between the SMs and neighborhood gateway (NG) based on the proposed PUF. The provable security analysis shows that not only the proposed protocol can stand secure in the Canetti–Krawczyk (CK) adversary model but also provides additional security features. Also, the performance evaluation demonstrates the significant improvement of the proposed scheme in comparison with the state-of-the-art.

## I. INTRODUCTION

SINCE the concept of the smart grid is based on using information and communication technology (ICT) in order to enhance the power grid in terms of safety, efficiency, and reliability, security against adversarial attacks is one of its major challenges. ICT provides communication alongside the electrical flow where a neighborhood gateway ($NG$) is able to gather ongoing updates from the smart meters ($SM$s) and send the control messages to them [1],[2]. Although ICT provides interesting features for the smart grid, it causes serious challenges in terms of security. The term security in the smart grid is often known by studying the impact of or dealing with several known attacks existing in a certain threat model. In most cases, an attacker can intercept the messages being transmitted through the communication link to acquire vital data, disrupt the communication, alter the transmitted data, or impersonate the $SM$ or $NG$. Additionally, an attacker can physically access to $SM$s' memory, obtain the stored secrets e.g. the cryptographic key, and break the security of the entire system [3]-[5].

### A. Related Works

Through recent years, many security protocols have been proposed to establish secure communication between the $SM$s and $NG$ in the smart grid. In 2014, Nicanfar et al. [6] presented an authentication and key management mechanism for smart grid, by updating the public, private, and multi-casting keys using a key generator. However, authors in [7] showed that this protocol is not suitable in practice. Tsai et al. [8] proposed an authentication protocol using bilinear pairings for smart grid communication. In 2018, the authors of [9] showed that Tsai et al.'s scheme has some security defects, hence, they presented another protocol based on bilinear pairings with better security properties. However, both of these schemes suffer from high computational overhead due to using bilinear pairings. In addition, according to Chen et al. [10], the scheme introduced in [9] has a security defect, where the $SM$s can be tracked and impersonated. In 2018, Abbasinezhad-Mood et al. [11] proposed an anonymous key agreement scheme for smart grids using elliptic curve cryptography (ECC). However, as shown by Braeken et al. in [12], their scheme lacks forward secrecy. Other public-key-based protocols have been proposed for smart grids in recent years. The authors in [13] presented a privacy-preserving authentication scheme using homomorphic encryption and bloom filter. Wazid et al. [14] proposed a three-factor authentication scheme based on fuzzy extractors and biometric properties. Although their protocol supports strong security and anonymity and provides dynamic addition of smart meters, still lacks efficiency due to using ECC. Qi et al. [15] proposed a two-pass privacy-aware key agreement protocol for the smart grid by using ECC. Mahmood et al. [16] proposed a protocol based on ECC and one-way hash function for the smart grid. However, the authors in [17], showed that the presented protocol in [16] is not able to provide session-key security, and is vulnerable to impersonation attacks. Uludag et al. [18] proposed a data collection approach for a smart grid using Rivest Shamir Adleman (RSA). They mainly presented some solutions for minimizing the data collection time in their protocol. Recently, Abbasinezhad-Mood et al. [19] proposed a privacy-preserving key agreement protocol based on ECC. They used Random-Oracle Model to evaluate the security of their protocol. The mentioned public-key-based schemes put lots of computational overhead on resource-constrained $SM$s. They are also vulnerable to physical attacks due to storing the long-term session key in $SM$s' memory.

[1] Department of Information and Communication Engineering, Aalto University, Espoo, Finland.
[2] Department of Electrical Engineering, Iran University of Science and Technology, Tehran, Iran.
[3] ETSI de Telecomunicación, Universidad Politécnica de Madrid, Madrid, Spain (corresponding author: diego.martin.de.andres@upm.es).
[4] Department of Computer Science and Engineering, University of South Florida, Tampa, USA.

At the same time, efforts have been stepped up to introduce more lightweight approaches to transferring information at each stage of authentication between $NG$ and $SM$. Zhang et al. [20] introduced a key establishment protocol based on hash function and symmetric algorithm for smart grid. However, their scheme imposes high data overhead for withstanding desynchronization attacks. Li et al. [21] proposed a Merkle hash tree-based authentication where $NG$ and $SM$ can securely communicate using the advanced encryption standard (AES). Liu et al. [22] proposed an authentication protocol based on the Lagrange polynomial (LP) for smart grids. They proved that their protocol has a better performance than Li et al.'s [21] scheme in communication, computational, and storage, cost. Although Abbasinezhad-Mood et al. [23] have presented a lightweight protocol for the smart grid by using pseudo-random number generation (PRNG) and the one-way hash function, their scheme lacks providing some important security features. The authors in [24] introduced a lightweight authentication scheme using XOR operation and the one-way hash function for smart grids. It is worth noting that no well-accepted adversary model was considered in [20]-[24] nor they are robust against physical attacks.

Although most of the mentioned schemes have certain flaws in both terms of security and efficiency, the same important drawback of these schemes is their vulnerability against physical attacks. Because the proposed schemes in [6]-[24] utilize the traditional key agreement protocols, which leads to storing of the longtime session key, the security of their scheme will be easily broken when an attacker reads the $SM$'s memory. Thus, physically unclonable functions (PUFs) have been introduced as an essential part of cryptographic protocols to prevent such physical attacks [25]-[30].

PUFs are defined as unique physical fingerprints of electronic devices. PUFs are particularly considered functions that map a set of challenges to a set of responses based on the intricate physical structure of its used circuit. In other words, the performance of PUFs completely depends on the uncontrollable variations in the integrated circuit (IC) manufacturing process it is practically hard to duplicate them even by having the exact manufacturing formula. The uncontrollable IC manufacturing process leads to generating random and unpredictable responses from PUFs which can be used for secure key generation. More details of the principles and definitions of different kinds of PUFs like Arbiter PUFs and RO PUFs can be found in [31],[32].

Gope et al. proposed an authentication scheme for secure communications based on reconfigurable PUFs (RPUFs) [25]. By using RPUFs, they could address the modeling attacks problem on strong PUFs. However, the utilization of RPUFs

TABLE I
ADVANTAGES AND DISADVANTAGES OF THE RELATED WORKS

| Scheme | Main Primitive | Advantage | Disadvantage |
|---|---|---|---|
| [6]-[19] | Public-key-based: bilinear pairings, ECC, RSA, CCM | $A_1$ | $D_2, D_3$ |
| [20]-[24] | Non-public-key-based: AES, SHA-256, XOR, PRNG | $A_2$ | $D_1, D_3$ |
| [25]-[30] | PUF-based: PUF, SHA-256, fuzzy extractors | $A_1, A_3$ | $D_2, D_4$ |
| Ours | PUF, SHA-256, PRNG, XOR | $A_1, A_2, A_3, A_4$ | -- |

CCM: Chebyshev Chaotic Map. $A_1$: CK-adversary model is considered. $A_2$: lightweight design. $A_3$: resistance against physical attack. $A_4$: efficient error correction technique for PUF. $D_1$: DY-adversary model is considered. $D_2$: inefficient design. $D_3$: vulnerable to physical attack. $D_4$: inefficient error correction technique for PUF.

leads to having an impractical protocol due to their limited number of supporting responses. Kaveh et al. [26] proposed an authentication scheme for the smart grid. Although their protocol could provide efficiency with physical security, they did not consider the PUF reliability issue. Ameri et al. [27] proposed a provably secure broadcast authentication protocol based on RO PUF and bloom filter. They used fuzzy extractors for PUF reliability based on Bose, Chaudhuri, and Hocquenghem (BCH) algorithm. However, their scheme puts a considerable computational burden on $SM$. Gope et al. [28] proposed a PUF-based key agreement scheme for the smart grid. They used a fuzzy extractor (BCH) for ensuring PUF reliability. Although they showed that their protocol is robust against physical attack, using heavyweight primitives led to up surging of the computational cost. Recently in 2020, Prosanta Gope [29] proposed a PUF-based authentication scheme. He has used the advantage of the BCH-based reverse-fuzzy extractor to ensure PUF reliability. Although he improved the former schemes which were based on fuzzy extractors, his protocol still lacks efficiency. Authors in [30] proposed an authentication scheme based on reliable PUFs by using the BCH error correction codes. They moved the weight of the decoding process toward $NG$ but still left a considerable amount of computational overhead on $SM$s. Table I shows how the previous schemes have been able to meet the important requirements of the smart grid, and what research gaps we aim to address in this paper.

In addition to fuzzy extractors, other methods have been developed in recent years for improving PUF reliability. Majzoobi et al. [33] proposed methods for majority voting of responses and classifying the challenges to different reliability sets in order to increase the stability of the corresponding responses in different environmental situations. However, it is shown that the majority voting-based reliability enhancement schemes are not sufficient due to their need to be executed repeatedly many times and store intermediate voting counts in volatile memory [34],[35]. Reconfigurable PUFs have also been proposed not only for enhancing PUF security but also for improving its reliability. However, these PUFs cannot be directly used for authentication purposes due to their ability to correct only a limited number of bit errors in their responses [36].

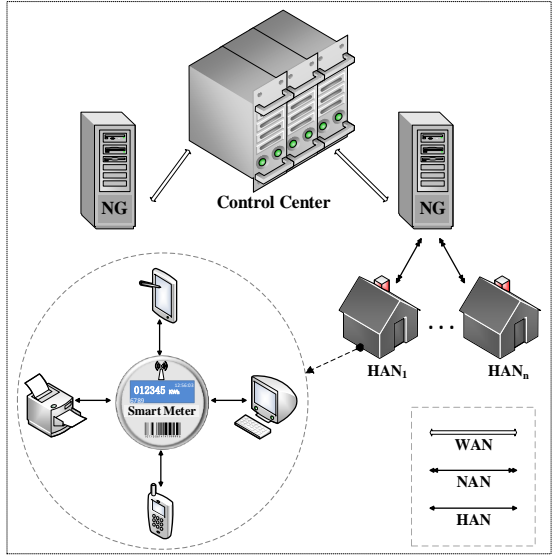

Fig. 1. The smart grid communication model.

TABLE II
NOTATIONS AND THEIR MEANINGS

| Symbol | Description |
| --- | --- |
| $h(.)$ | One-way hash function |
| $SM_j$ | $j^{th}$ smart meter |
| $i$ | $i^{th}$ time interval of data transmission |
| $ID_i^j$ | Identifier of $SM_j$ in $i^{th}$ authentication |
| ms | Millisecond |
| B | Byte |
| $\oplus$ | The XOR operator |
| $\parallel$ | The concatenation operator |
| $\|X\|$ | The size of parameter $X$ |
| °C | Degrees Celsius |

## B. Motivation

Although the PUF-based protocols presented in [25]-[30] have addressed the vulnerability to physical attack of schemes proposed in [6]-[24], the lack of finding proper solutions for PUF reliability and designing the protocol in a lightweight manner is their most important disadvantage. Most of PUF-based protocols have proposed fuzzy extractor to be used for generating robust PUF response which not only imposes high computational cost to the system but also has its own security defects [37]-[39]. Furthermore, in most of the previous schemes, either the threat model is not considered or the Dolev-Yao (DY) adversary model [40] is used while it is well-established that the Canetti–Krawczyk (CK) adversary model [41] can provide a more comprehensive threat model for the smart grid environment. All these issues have motivated us to propose a PUF-based lightweight authentication protocol that not only is secure in the CK-adversary model but also optimally uses robust PUF responses to securely generate cryptographic keys.

## C. Our Contribution

The contributions of this paper are as follows:

- We propose an on-chip-error-correcting (OCEC) PUF to efficiently collect reliable bits to generate robust keys.
- We propose a lightweight authentication protocol based on OCECPUF that provides some key features such as mutual authentication, data confidentiality and integrity, $SM$'s anonymity and privacy, forward secrecy, and physical security in the presence of a CK-adversary.
- We provide a formal security analysis to prove that the proposed scheme is secure in presence of probabilistic polynomial time (PPT) adversaries.
- By eliminating fuzzy extractors and only utilization of lightweight cryptographic primitives like hash functions, our OCECPUF-based protocol dramatically outperforms the previous works in terms of computational cost.

The remainder of this paper is organized as follows. The system/threat model is discussed in section II. Section III elaborates OCECPUF design and the proposed protocol. The formal and informal security and performance analysis are shown in sections IV, V, and VI, respectively, and section VII draws the conclusion of this paper.

## II. SYSTEM AND THREAT MODEL

The communication layer of the smart grid consists of three levels. At the first level, a home area network (HAN) includes several smart appliances and a $SM$ to collect the electricity data. Each $SM$ is a resource-constrained device that has a PUF. Since the proposed PUF is embedded in the $SM$s, any physical attempts on removing the PUF from it led to the destruction of the $SM$ and its corresponding PUF. In the second level, the neighborhood area network (NAN) consists of one $NG$ that gathers the updates from a certain number of $SM$s and sends the control signals back to them. In the final level called the wide area network (WAN), the control entity supervises a small number of $NG$s in the top level and makes the final decisions. Fig. 1 shows the smart grid communication layer, where in we concentrate on the communications between the $SM$s and $NG$ in this paper.

For this paper, we consider the widely accepted CK model for adversaries. In the CK-adversary model, which is a stricter model than the DY model, not only do adversaries have full access to the communication link but also, they can physically capture involved parties of the protocol. In other words, an adversary may eavesdrop, modify, try to decrypt, and replay the messages, which are transmitted in the communication link or disrupt the network by performing a denial of service (DoS) attack. He/she can also impersonate both $SM$ and $NG$. In addition, an adversary is able to capture an $SM$ and tries to obtain the important secrets stored in its memory. In addition, the outcome of ephemeral secrets leakage of one session will also be considered in the CK model. As per the system and threat model, the objective of this paper is to suggest a PUF-based lightweight authentication scheme that can withstand all potential attacks in the NAN communication network.

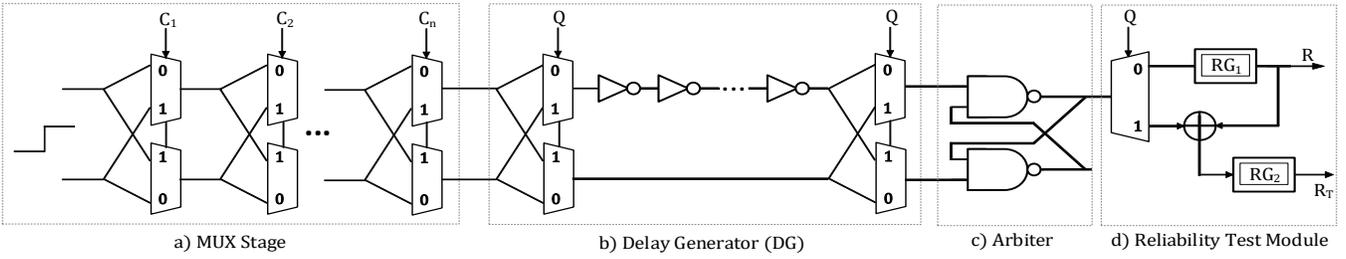

Fig. 2. The general schematic of the proposed PUF model.

## III. THE PROPOSED PUF-BASED PROTOCOL

In this section, we first elaborate the proposed used PUF model embedded in $SM$s. After that, we elaborate our PUF-based scheme. Table II shows all used notations in this paper and their meanings.

### A. The Proposed PUF Model

The random digits of PUFs must meet the necessary requirements to be suitable for usage in cryptographic applications. One important property of PUFs is their reproducibility or reliability. This property states that when a challenge $C$ is input to a PUF instance in different situations, the generated responses (for the same challenge) should be identical. However, there are always some errors in PUF responses in real implementations due to environmental factors like temperature variations. In this paper, in order to be able to use a robust PUF in the authentication process between the $NG$ and $SM$, we use an efficient PUF model including an arbiter PUF with a simple error detection circuit. The key idea of this part is to avoid any preprocessing on PUF responses like error correction code-based solutions which significantly increase the hardware and computational costs. Instead, we aim to optimally remove the errors at the chip level.

Arbiter PUF is the most famous PUF that is generally based on the two-stage multiplexers (MUXs) which are controlled by the challenge bits and subsequently generate two signals with different delay-time. These signals will be input to an arbiter which generates a 0 or 1 digit based on which signal arrives first, i.e. when the delay of the signal from the upper path $d_1$ is bigger than the delay of one from the lower path $d_2$ ($\Delta d = d_1 - d_2 > 0$), the arbiter generates 0, otherwise, it generates 1. Fig. 2 demonstrates the proposed PUF structure in this paper which consists of four main modules including MUX stages, delay generator ($DG$), Arbiter, and reliability test module ($RTM$). According to Fig. 2(a), the signals from the upper and lower paths with delays $d_1$ and $d_2$ enter the $DG$ after passing the last MUXs in the MUX stage. $DG$ consists of one delay-line including some NOT gates, and two 2-2 MUXs controlled by signal $Q$. When a challenge is input, the control signal $Q$ is generated by a controller (simply by a state machine) to control the PUF operations which takes the both values of 0 and 1. First, considers that $Q = 0$, which connects the delay-line to the upper path. Now, by assuming the delay of the delay-line is to be $D$, the delay of the upper path will be $d_1 + D$. Therefore, $d_1 + D - d_2 = \Delta d + D$ is fed to the arbiter. Secondly, assume that $Q = 1$, which connects the delay-line to the lower path. In this case, the delay of the lower path will be $d_2 + D$, and $d_1 - (d_2 + D) = \Delta d - D$ is fed to the arbiter. Fig. 2(b) depicts the $DG$ architecture.

The core idea of adding a delay $D$ to the upper and lower paths of the Arbiter PUF is to recognize paths that their delay difference is less than a certain value ($\Delta d < D$), because the errors occurrence in the final PUF responses is due to the passing of the signal through such paths. In other words, if we define a threshold $D$ large enough, we can recognize the stable paths with most probability that would remain stable even in a very unstable environment. What comes in the following, shows that how a desire value for $D$ can be obtained.

For a typical Arbiter PUF, a flip-flop arbiter is often used to determine the faster/slower delay paths. It has been shown that employing flip-flops as arbiters causes a 10% deviation in the routing path [42]. Therefore, we use the cross-coupled NAND gates for arbitration as depicted in Fig. 2(c).

After the signals pass through $DG$, Arbiter delivers the response bit to $RTM$. As shown in Fig. 2(d), $RTM$ consists of a 1-2 demultiplexer (DMUX) controlled by $Q$, an exclusive-OR (XOR) gate, and two registers ($RG_1$ and $RG_2$). First, when $Q = 0$, the generated response $R$, acquired by entering $\Delta d + D$ the Arbiter, is stored in $RG_1$. Then when $Q$ is set to 1, the new response $R'$, acquired by entering $\Delta d - D$ the Arbiter, is fed together with $R$ to the XOR gate and generates a reliability test bit, $R_T$ ($R_T = R \oplus R'$), which is stored in $RG_2$. Now, we explain how $R_T$ can detect unstable digits.

First, assume that the delay of signal from the upper path is bigger than the one from the lower path ($\Delta d = d_1 - d_2 > 0$) and the corresponding response will be 0. When $Q = 0$, as mentioned before, $DG$ adds the delay line to the signal from the upper path and $\Delta d + D$ is fed to the arbiter. Since $D$ is a positive value and $\Delta d + D > 0$, Arbiter generates the response 0 which is stored in $RG_1$ through DMUX. When $Q$ switches to 1 for the same challenge, $DG$ adds the delay line to the lower path and $\Delta d - D$ is fed to the Arbiter. Now, we consider two important cases. 1) If $\Delta d > D$, then $\Delta d - D > 0$ and Arbiter generates 0 again. Since the two generated responses are the same, the output of XOR operation ($R_T = R \oplus R'$) is 0 and this 0 will be stored in $RG_2$ which represents the reliability test bit. $R_T = 0$ states that the difference between the selected paths is greater than a desired value ($D$) and therefore the generated response is stable. 2) If $\Delta d < D$ then $\Delta d - D < 0$ and Arbiter generates 1. Now, $R_T = 1 \oplus 0 = 1$ will be stored in $RG_2$ representing that the delays of the upper and lower paths are too close to each other and the generated bit is prone to error in the PUF response. In other words, the value of $R_T$ represents the stability of the generated digit i.e. $R_T = 0$ states that the generated response is stable, and $R_T = 1$ states that the generated digit is unstable.

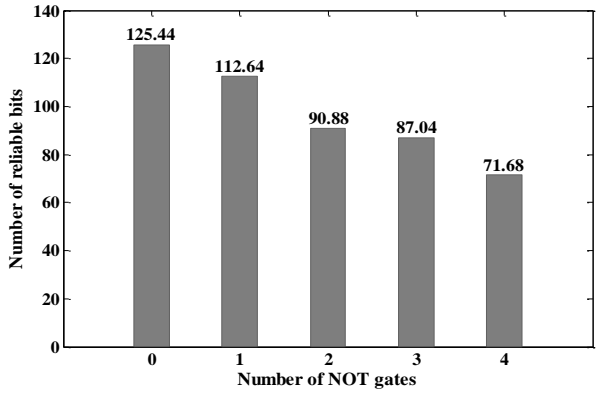

Fig. 3. Average selected robust bits per each number of used NOT gates.

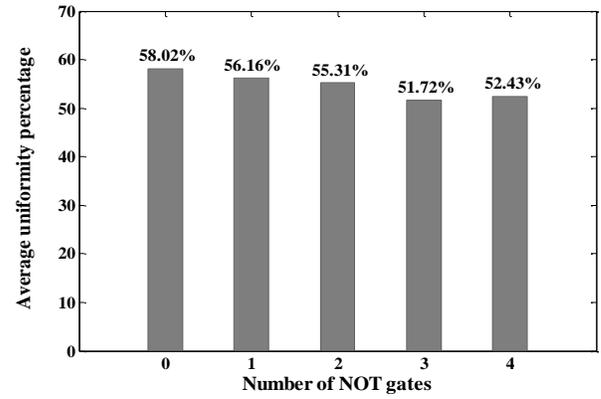

Fig. 5. Average uniformity for different number of used NOT gates.

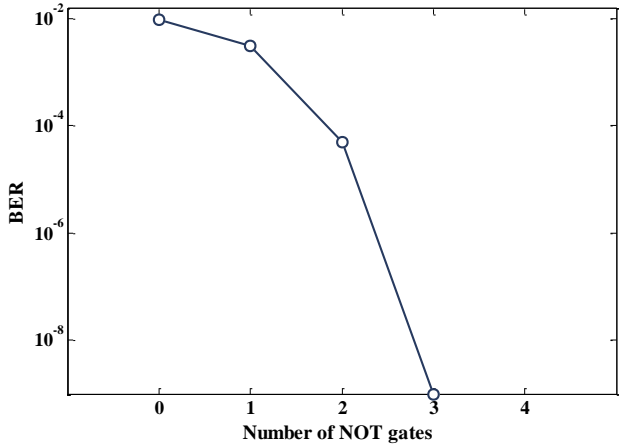

Fig. 4. BER for different number of used NOT gates.

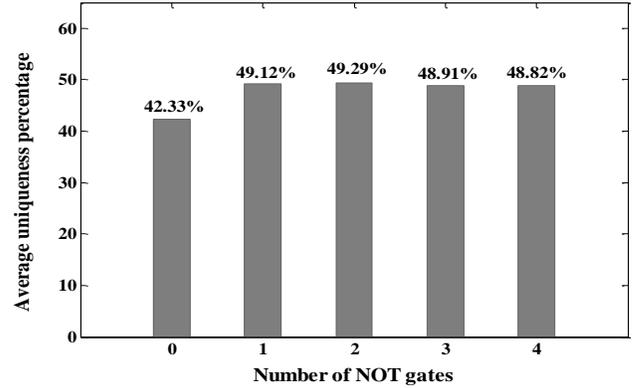

Fig. 6. Average uniqueness for different number of used NOT gates.

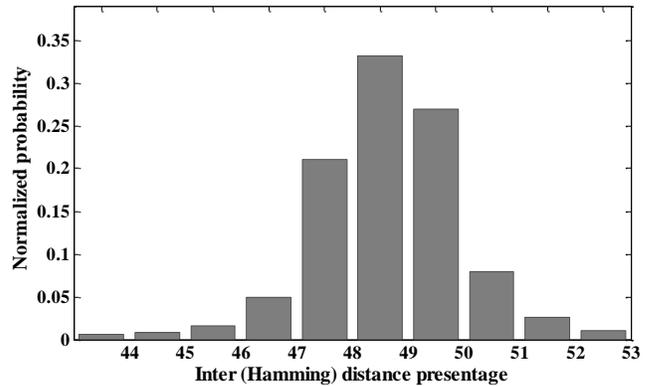

Fig. 7. Inter Hamming distance histogram for 4 NOT gates in the delay-line.

A similar analysis comes when the delay of the signal from the lower path is bigger than the one from the upper path ($\Delta d = d_1 - d_2 < 0$). Here, when $|\Delta d| > D$, then $R_T = 0$ and it is concluded that the generated response stored in $RG_1$ is stable. But when $|\Delta d| < D$, $R_T = 1$ and the generated digit is an unstable response bit. It is worth noting that only responses with $R_T = 0$ can be used for cryptographic purposes, while those with $R_T = 1$ are simply discarded.

In order to determine the proper number of NOT gates in the delay line (proper $D$), we implement forty 128-bit 32-stage proposed PUF models on ten FPGA boards from Spartan-6 family named Xilinx XC6SLX45, which is known as a low-cost and public FPGA. We consider various numbers of NOT gates in the delay line from zero to four and collect the stable response bits for each case. The stable responses are the ones in which their reliability test bit stored in $RG_2$ is 0 ($R_T = 0$). To that end, we first evaluate our PUFs under normal temperature (25°C) to find the stable bits for various numbers of NOT gates in $DG$. We also analyze PUF uniformity and uniqueness under normal temperature. After then, we evaluate the PUFs with their selected reliable responses under different temperatures (0°C–80°C) to compute the final bit error rate (BER) of our PUF model per different NOT gates in $DG$. Fig. 3 shows the average selected robust responses per each number of used NOT gates. Furthermore, Fig. 4 demonstrates the BER for different used NOT gates. According to this figure, BER exponentially decreases by increasing the number of used NOT gates in the delay-line so that with four NOT gates in the delay-line, we had no error under different temperatures (0°C–80°C). Fig. 5 shows the average uniformity of PUF responses for different number of used NOT gates from 0 to 4. 52.43% bias has been achieved for 4 NOT gates

TABLE III
PERFORMANCE COMPARISON WITH OTHER PUF RELIABILITY-ENHANCEMENT SOLUTIONS

| Scheme | BER | Used LUT | Used FF |
|---|---|---|---|
| DSC PUF [43] | $6.14 \times 10^{-9}$ | 251* | 235* |
| 4-DFF MA-PUF [44] | 0.16% | 302 | 512 |
| SR latch MA-PUF [44] | 0.17% | 506 | 256 |
| BST-APUF [45] | $< 10^{-9}$ | 150 | 47 |
| R-PUF [46] | 6.7% | 1536 | 128 |
| TQR-APUF [47] | 0.577% | 192 | 132 |
| This work | $< 10^{-9}$ | 104 | 38 |

*The overhead of the PUF is not included.

in the delay line, almost close to ideal value, while the number of bit errors in this case is zero at the different environmental situations. The uniqueness property of the proposed PUF can be analyzed using Fig. 6 and 7. In the same case (4 NOT gates in the delay-line), 48.82% uniqueness has been achieved with a near-to-ideal inter Hamming distance distribution. It is worth mentioning that since the number of reliable bits was different for every experiment, we considered the minimum bit length for measuring the inter Hamming distance in each case.

As a brief conclusion, we have achieved fully reliable and close-to-ideal response bits for our design that can be directly used for authentication and key generation purposes without the need of employing any error correction codes. This leads to significant optimization in hardware area and computational cost. Table III shows the hardware overhead and BER of the proposed PUF in comparison with other reliable PUF designs proposed in [43]-[47]. According to this table, not only does our design has lower BER but also outperforms the state-of-the-art in term of hardware overhead. Now, we describe our protocol based on OCECPUF in the following subsections.

*B. Registration Phase*

At this stage, $SM_j$ registers itself to $NG$ as follows. At first $SM_j$ sends a request to $NG$ to be registered. $NG$ investigates the authenticity of $SM_j$ and then computes two values $ID_j^0$ and $C_j^0$ as the first identity and challenge bits, respectively. Then $SM_j$ stores $ID_j^0$ and input $C_j^0$ to its OCECPUF to generate the corresponding response, i.e., $R_j^0 = OCECPUF_j(C_j^0)$. Then it computes $K_j^0 = h(R_j^0)$ and sends $K_j^0$ to $NG$, removes both $C_j^0$, $R_j^0$ from its memory and stores $ID_j^0$ instead. On the other side, $NG$ stores $ID_j^0$, $C_j^0$, and $K_j^0$ in the raw of data base that is related to $SM_j$. Based on the system policies, this phase can be repeated over very long predefined periods. It is also worth noting that all the communications in the registration phase are done in a secure channel.

*C. Communication Phase*

With no loss of the general understanding, we only consider the i-th communication between $NG$ and $SM_j$, which can be extended to other components at different time intervals. First, $SM_j$ generates a nonce $N_j^i$ and sends it alongside its identity $ID_j^i$ to $NG$ for starting the authentication process. $NG$ then locates $ID_j^i$ in its database and finds the corresponding pair $(C_i^j, K_i^j)$. Then, it generates a pseudo-random number $r_{NG}^i$ and computes a parameter as $E_{NG}^i = ((M_{NG}^i \oplus r_{NG}^i) \parallel r_{NG}^i) \oplus K_j^i$ where $M_{NG}^i$ is the control message for $SM_j$. After that, $NG$ computes a verifier as $V_{NG}^i = h(K_j^i, C_j^i, M_{NG}^i, r_{NG}^i, N_j^i)$ and then sends the packet $\{C_j^i, V_{NG}^i, E_{NG}^i\}$ to $SM_j$.

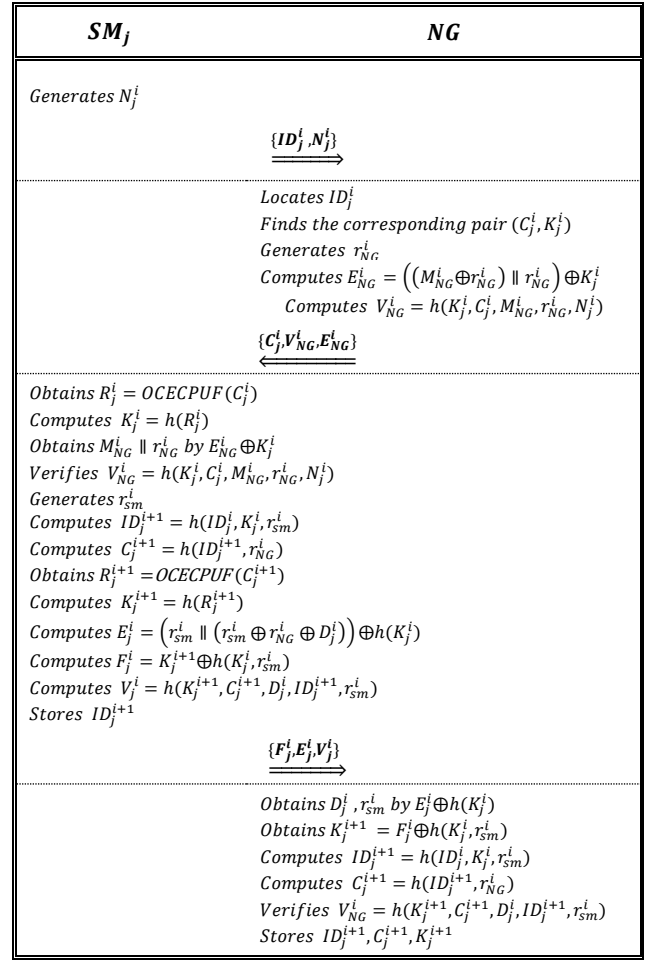

Fig. 8. The proposed PUF-based protocol.

After receiving the packet, $SM_j$ executes its PUF by using $C_j^i$ as $R_j^i = OCECPUF(C_j^i)$ and uses the reliable response $R_j^i$ to compute $K_i^j$ as $K_j^i = h(R_j^i)$. Now $SM_j$ can decrypt $E_{NG}^i$ to obtain $M_{NG}^i$ and $r_{NG}^i$ by computing $E_{NG}^i \oplus K_j^i$ and verify $V_{NG}^i$. If $V_{NG}^i$ passes the verification, $SM_j$ generates a pseudo-random number $r_{sm}^i$ and computes the values $ID_j^{i+1} = h(ID_j^i, K_j^i, r_{sm}^i)$, $C_j^{i+1} = h(ID_j^{i+1}, r_{NG}^i)$ and $R_j^{i+1} = OCECPUF(C_j^{i+1})$ as its new ID, new challenge, and new response, respectively. Then, it computes the new key $K_j^{i+1} = h(R_j^{i+1})$ for the next authentication and encrypts its new key, random number, and data report ($D_j^i$) by computing $F_j^i = K_j^{i+1} \oplus h(K_j^i, r_{sm}^i)$ and

$E_j^i = \left(r_{sm}^i \parallel \left(r_{sm}^i \oplus r_{NG}^i \oplus D_j^i\right)\right) \oplus h(K_j^i)$. $SM_j$ also computes a verifier as $V_j^i = h(K_j^{i+1}, C_j^{i+1}, D_j^i, ID_j^{i+1}, r_{sm}^i)$ and sends $\{F_j^i, E_j^i, V_j^i\}$ to $NG$. At the end, $SM_j$ stores $ID_j^{i+1}$ in its memory and removes all other parameters. After receiving the packet, $NG$ obtains $r_{sm}^i$ and $D_j^i$ by computing $E_j^i \oplus h(K_j^i)$, and then it obtains $K_j^{i+1}$ by computing $F_j^i \oplus h(K_j^i, r_{sm}^i)$. After that, $NG$ computes $ID_j^{i+1} = h(ID_j^i, K_j^i, r_{sm}^i)$ and $C_j^{i+1} = h(ID_j^{i+1}, r_{NG}^i)$ and then verifies $V_j^i$. If $V_j^i$ passes the verification process, the mutual authentication is done, $NG$ accepts the message, and replaces $ID_j^{i+1}$, $C_j^{i+1}$, and $K_j^{i+1}$ with the previous ones in its database. Fig. 8 depicts our protocol.

## IV. FORMAL SECURITY ANALYSIS

In this section, we aim to prove the security of the proposed scheme in the presence of PPT adversaries. It is obvious that the authenticity of the scheme only relies on using a proper cryptographic hash function. As a result, because the one-way hash function $h$ is collision-free for PPT computers, we suffice to analyze the authenticity of the proposed scheme only informally. Thus, in what comes next we aim to prove the confidentiality of the scheme [48]. Furthermore, we provided a formal security verification using ProVerif and an informal security analysis in a supplementary file. We showed that the proposed protocol can resist message analysis, alteration, impersonation, replay, ephemeral secret leakage DoS, and physical attacks, and provides privacy, anonymity, and forward secrecy.

In order to set up the proof first we introduce two lemmas which will be useful for providing the main proof.

***Lemma 1***: the function $OCECPUF(.)$ is a true random number generator and the resulted value $R_i = OCECPUF(c_i)$, where $c_i$ is an arbitrary challenge, is indistinguishable from a same length string created by a pseudo-random generator for any PPT adversary.

***Proof***: look into the randomness property of the OCECPUF.

***Lemma 2***: the one-way collision-free hash function $h(.)$ with a random input and uniform distribution is computationally indistinguishable from a same-length string created by a pseudo-random generator.

***Proof:*** in order to prove this lemma first we introduce a game. In this game at first, two random numbers $x$ and $r$ are generated. Then, a coin is flipped, and based on which side it lands, the parameter $u$, which is either $h(x)$ or $r$, will be given to the distinguisher $D$. Now, the distinguisher $D$'s goal is to find out that the given string $u$ is a hashed value created by the hash function $h(.)$ or it is a random string generated by a pseudo-random number generator.

Hence, we have to prove that the probability of $D$'s success is negligible more than a pure guess. In other words, we have to show that the following equation holds:

$$\left|pr[D^{u=h(x)} = 1] - pr[D^{u=r} = 1]\right| \leq negl\ (n)$$

According to the definition of pseudo-random number generators, it is obvious that when the string is created from a pseudo-random number generator $r \leftarrow g(s)$, the probability of success for the distinguisher adversary $D$ is equal to a pure guess which means $pr[D^{u=r} = 1] = \frac{1}{2}$. Now in the case where $u$ is a hashed value $h(x)$ where $x$ is chosen randomly and uniformly, the probability of $D$'s success is:

$$pr[D^{u=h(x)} = 1] = \frac{1}{2} + \epsilon(n)$$

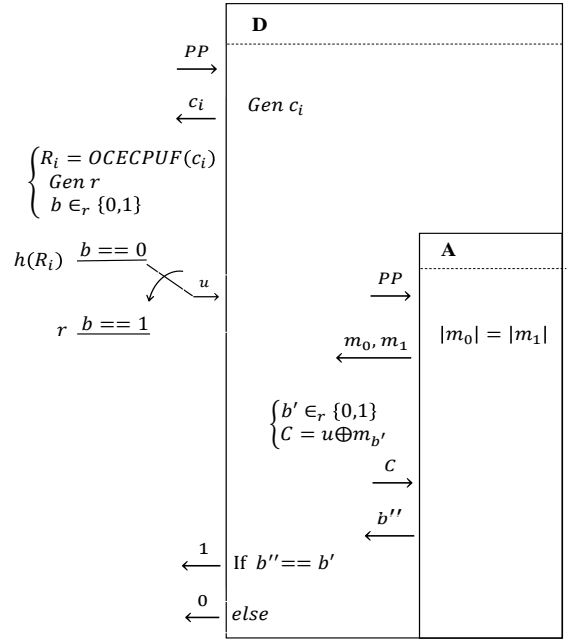

Fig. 9. The adversaries model in the provided game.

Where the $\frac{1}{2}$ comes from the pure guess and all the additional information that $D$ can achieve from $h(x)$ with an unknown and random $x$ is shown by $\epsilon(n)$. Hence, we have:

$$\begin{aligned}\epsilon(n) &= pr[find\ x'|h(x') = u] = pr[find\ x'|h(x') = h(x)]\\&= pr[find\ x'|h(x') = h(x), x' \neq x]\, pr[x' \neq x] + pr[find\ x'|h(x') = h(x), x' = x]\, pr[x' = x]\end{aligned}$$

As $x$ is chosen randomly, the probability $pr[x' = x] = \frac{1}{2^{l(n)}} = negl\ (n)$ where $l(n)$ is the length of the message. Also, the probability $pr[find\ x'|h(x') = h(x), x' \neq x]$ is equivalent to finding a collision in the hash function that is negligible. As a result:

$$\epsilon(n) = negl(n) \times \left(1 - \frac{1}{2^{l(n)}}\right) + pr[find\ x'|h(x') = h(x), x' = x] \times negl\ (n) \leq negl\ (n)$$

Hence:

$$pr[D^{u=h(x)} = 1] \leq \frac{1}{2} + negl\ (n)$$

Which means:

$$\left|pr[D^{u=h(x)} = 1] - pr[D^{u=r} = 1]\right| = \left|\frac{1}{2} + negl\ (n) - \frac{1}{2}\right| \leq negl\ (n)$$

And the lemma is proved. In other words, this lemma presents that if a PPT distinguisher $D$ exists that is able to distinguish between the output of a one-way collision-free hash function and a random string, then another adversary exists which is able to find a collision in the hash function.

***Theorem***: the proposed scheme has indistinguishable encryption security in the presence of a PPT eavesdropper adversary.

This theorem implies that any PPT eavesdropper cannot make difference and link an encrypted message to its corresponding plain text. We bring the proof for $E_{NG}$ and the same analysis can be brought for $E_j^i$. First, we name the proposed scheme $\pi$ as below:

$$\pi: \begin{cases} Gen & k \leftarrow \{0,1\}^n \\ Enc & C = m \oplus G(k) \\ Dec & m = C \oplus G(k) \end{cases}$$

Where $k$ is the challenge, $G(k) = h(OCECPUF(k))$ is the key, $C = E_{NG}^i$ and $m = (M_{NG}^i \oplus r_{NG}^i) \parallel r_{NG}^i$.

Now, in order to setup the proof we introduce a new game depicted in Fig. 9. As it can be seen from Fig. 9 the game includes two PPT adversaries. The adversary $A$ is the one wanting to attack the proposed scheme and the distinguisher $D$ is another adversary which uses the information of $A$ to distinguish between two random strings.

After both $A$ and $D$ receive the public parameters $PP$ which includes $\{l(n), h(.)\}$, the game starts with $A$ choosing two arbitrary messages $m_0, m_1$ with the same length and sending them to $D$. $D$ sends a challenge $c_i$ to the oracle. The oracle flips a coin and based on which side it lands, sends $h(OCECPUF(c_i))$ or a random string $r$ generated by a pseudo random generator. Then, $D$ flips a coin and encrypts either of $m_0$ or $m_1$ and sends it to $A$. Now, adversary $A$ succeeds if it is able to find $b'$ with probability more than pure guess $\frac{1}{2}$.

We prove that if an adversary $A$ exists which is able to distinguish $C$ and finds $b'$ with probability $\epsilon$ which is not negligible, then adversary $D$ can distinguish between a hashed value, with a random and unknown input, and a random string created by a pseudo random generator which is in contradiction with lemma 2. At first, consider the case where $b == 1$ and thus, $u = r$. $D$ choses $b' \in_r \{0,1\}$ and encrypts the message $m_{b'}$ by computing $C = r \oplus m_{b'}$ and gives it to $A$. Now, the goal of $A$ is to determine $b'$ to finds out which of the $m_i$ is encrypted. In this case, the security of the scheme is equal to an one-time pad crypto system which is believed to have a perfect security. Hence, the probability of success for the PPT adversary $A$ in guessing which of the messages is encrypted is equal to pure guess.

$$pr[Privk_{A,\pi}^{eav}(n,r) = 1] = \frac{1}{2}$$

Now consider the case where $b == 0$ hence $u = h(R_i)$ and the encrypted message will be $C = h(R_i) \oplus m_{b'}$. In this case we consider what happens if the adversary $A$ can succeed and finds $b'$ with probability $\frac{1}{2} + \epsilon(n)$, where $\epsilon(n)$ is not negligible. Thus, we consider:

$$pr[Privk_{A,\pi}^{eav}(n, h(R_i)) = 1] = \frac{1}{2} + \epsilon(n)$$

It is worth noting that, the adversary $A$ attacks the proposed scheme directly and the probability of its success means the security breakdown of the proposed scheme. Now, we compute the probability of $D$'s success which is its ability to distinguish between $h(R_i)$ and $r$. Hence:

$$|pr[D^{h(R_i)} = 1] - pr[D^r = 1]|$$
$$= |pr[Privk_{A,\pi}^{eav}(n, h(R_i)) = 1]$$
$$- pr[Privk_{A,\pi}^{eav}(n,r) = 1]|$$
$$\leq \left|\frac{1}{2} + \epsilon(n) - \frac{1}{2}\right| = \epsilon(n)$$

Which contradicts the lemma 2. As a result, we proved that if an adversary like $A$ exists which can attack the proposed scheme, then a distinguisher $D$ exists which can distinguish between a random string and a hashed value with an unknown and random input and based on the lemma 2, if such $D$ exists then, another attacker exists which can find a collision in the hash function. We proved the indistinguishably encryption security of the proposed scheme on $SM$'s side. Similarly, the same analysis can be provided for the $NG$'s side.

## V. INFORMAL SECURITY DISCUSSION

In this section, we describe how the proposed protocol can stand against the possible attacks in the CK-adversary model.

### A. Resistance to Message Injection and Modification Attacks

In these types of attack scenarios, the attacker's objective is to alter or fabricate a message by pretending to be one of the authorized parties on the network, tricking the other party into believing the message is from a trustworthy source. In order to fully examine this attack method, because our system involves two-way communication, the attacker can impersonate both NG and SM. In either scenario, the attacker has access to the transmitted packets and their aim is to modify the message and create a corresponding verifier that will pass the verification process at the other end. However, because the hash function used in computing the verifier is one-way, the attack is not feasible. In other words, the polynomial time computer cannot compute valid verifiers as $V_{NG}^i = h(K_j^i, C_j^i, M_{NG}^i, r_{NG}^i, N_j^i)$ and $V_j^i = h(K_j^{i+1}, C_j^{i+1}, D_j^i, ID_j^{i+1}, r_{sm}^i)$ by having the transmitted packets only and without having $K_j^i, r_{NG}^i, K_j^{i+1}$ and $r_{sm}^i$. Thus, the message injection and modification attacks are not successful.

### B. Resistance to Replay Attack

In replay attacks, the adversary uses obsolete packets, which were transmitted in previous communications, and replays them in the network. However, as in our scheme, both NG and SMs generate new random numbers and nonce, and use them in the verifiers $V_j^i$ and $V_{NG}^i$ for each authentication phase, it will be detected if the adversary reuse the old packets.

Besides, the session keys are randomly created by a PUF and are updated per each communication; therefore, the probability of previous packets passing the verification is negligible.

*C. Resistance to Message Analysis Attack*

Message analysis attack issues confidentiality of message where the adversary tries to get access to the plaintext. Here, only $E_{NG}^i$, $E_j^i$, and $F_j^i$ are subjects to this attack. As seen in the proposed protocol, $E_{NG}^i$, $E_j^i$, and $F_j^i$ are encrypted with the session keys, which are secret, used only once, and updated in each communication run; hence, the security of our scheme will be reduced to the security of a one-time pad cryptosystem. Therefore, the message analysis attack is not possible for a PPT adversary.

*D. Resistance against DoS Attack*

This type of attack is designed to overwhelm the network to shut it down. The attacker is not trying to send meaningful messages or make them appear legitimate, but rather, their goal is to force the trusted parties on the network to consume extra computational resources. Here, the aim is to make the network parties preoccupied, causing them to miss receiving genuine messages from other authorized parties. Since our scheme facilitates communication in both directions, this type of DoS attack can be executed on both ends. Though, because of the high computational capacity of $NG$, we can ignore this attack on $NG$'s side and only consider the attack on $SM$. DoS attack on $SM$'s side can be done when $NG$ sends the packet $\{C_j^i, V_{NG}^i, E_{NG}^i\}$ to $SM_j$. To that end, the attacker sends spurious packets e.g., $\{X, Y, Z\}$ in form of $\{C_j^i, V_{NG}^i, E_{NG}^i\}$ to $SM_j$. However, the first thing $SM_j$ does is to run its PUF, and compute two hashes for generating $K_j^i$ and verifying $V_{NG}^i$, respectively. Therefore, by performing one PUF and two hash functions, which are insignificant, the packet will be ignored. It goes without saying that the probability of the random set $\{X, Y, Z\}$ passing the verification stage is negligible.

*E. SMs' Privacy and Anonymity*

In the proposed scheme, $SM_j$ updates the ID for every communication stage i.e., $SM_j$ uses its ID only once. As a result, this scheme supports a good level of privacy and anonymity for $SM_j$ against either the adversary or other trusted parties in the grid. In other words, only knows about $SM$s' activities. Therefore, our scheme in this paper provides the privacy and anonymity of $SM$s.

*F. Forward Secrecy*

Forward secrecy means that the revealing of one session key does not lead to the revealing of other past session keys. As seen in section III, the i-th and i+1-th session keys are computed as $K_j^i = h(R_j^i)$ and $K_j^{i+1} = h(R_j^{i+1})$, respectively, where $R_j^i$ and $R_j^{i+1}$ are two different PUF responses, hence, the session keys are computed completely independent to each other. In other words, having one session key would not give an adversary any useful information to compute the other session keys. In addition, as $F_j^i = K_j^{i+1} \oplus h(K_j^i, r_{sm}^i)$, by considering that the i+1-th session key, $K_j^{i+1}$, is compromised, the adversary can obtain the value $h(K_j^i, r_{sm}^i)$, however, in order to find the former session key $K_j^i$, the adversary has to find a collision in the one-way cryptographic hash function that is impractical for PPT computers.

*G. Ephemeral Secret Leakage Attack*

As mentioned earlier, in CK model, the short term secrets of an arbitrary session can be revealed, and the adversary's goal is to achieve as much as information it can from them and use them to attack the protocol in other sessions. In our scheme, the short-term secrets are the two random numbers $r_{NG}$ and $r_{sm}$. By knowing $r_{NG}$ the adversary can obtain the right half of the session key $K_j^i$ from $E_{NG}^i$, however, as the creation of the next key is independent to the previous key and is computed by running a PUF on a new challenge, the other session keys will not be revealed to the adversary. In addition, by knowing $r_{sm}$, the adversary can obtain right half of the $h(K_j^i)$, which does not reveal any additional information.

It is worth mentioning that as our scheme does not have a long-time secret key, the security of it strongly depends on the temporary session keys $K_j^i$. In other words, if $K_j^i$ is revealed to an adversary the next session keys can also be obtained while the former session keys cannot (forward secrecy). However, obtaining the session key is not possible for the adversaries as it is not stored in memory, and achieving it requires to run a PUF which requires to completely capture a smart meter, where at that point $NG$ will easily stop the communication with the captured $SM$. Furthermore, since revealing a session key will not expose the former session keys, the adversary cannot decrypt previous messages.

*H. Resistance to Physical Attack*

When examining physical attacks, only the non-volatile memory of the SMs is believed to be susceptible to being compromised by an attacker. This implies that the volatile memory, which is employed for keeping the confidential intermediary values throughout the communication process, will be erased once the communication session is completed and cannot be accessed by an adversary [49],[50]. As noted earlier, one of the goals of this paper was to make our scheme invulnerable to physical attacks by using the proposed SECPUF and eliminate the need of storing the keys in the memory. As a result, when $SM$s are captured by an adversary, he/she will not reveal any secret related to the session key, because the session keys are generated through performing SECPUF in each communication and there is no need for $SM$s for storing them. The only information the smart meters store in their memory is their pseudo-identity, however when the smart meter is captured by an adversary, knowing the pseudo-identity of it will be trivially observable. Note that, our scheme has forward secrecy for pseudo-identity, because the pseudo-identity is computed as $ID_j^{i+1} = h(ID_j^i, K_j^i, r_{sm}^i)$ hence, by knowing $ID_j^{i+1}$, computing former identities like $ID_j^i$ is equivalent to finding a collision in the hash function.

TABLE IV
COMPARATIVE COMMUNICATION OVERHEAD

| Scheme | [27] | [28] | [29] | [30] | Ours |
|---|---|---|---|---|---|
| Overhead | 112B | 162B | 170B | 224B | 192B |

TABLE V
EXECUTION TIME OF CRYPTOGRAPHIC OPERATIONS ON AT91SAM3X8E

| Cryptographic Operation | Execution Time |
|---|---|
| SHA-256 Hash Function | 39.2 $\mu s$ |
| AES-256 EBC Encryption | 198.48 $\mu s$ |
| AES-256 EBC Decryption | 309.67 $\mu s$ |
| Pseudo Random Number Generation | 82.3 $\mu s$ |
| Polynomial Generation | 3.6 $ms$ |
| BCH-based Secure Sketch | 2.27 $ms$ |
| BCH-based Secure Recover | 4.39 $ms$ |
| 128-bit Arbiter PUF | 160.7 $\mu s$ |
| 128-bit OCECPUF | 353.5 $\mu s$ |

TABLE VI
COMPARATIVE COMPUTATIONAL COST FOR ONE PROTOCOL EXECUTION

| Cost | [27] | [28] | [29] | [30] | Ours |
|---|---|---|---|---|---|
| $T_h$ | 6 | 7 | 7 | 5 | 8 |
| $T_{Enc}$ | × | × | × | × | × |
| $T_{Dec}$ | × | × | × | × | × |
| $T_{RNG}$ | × | 1 | 1 | × | 1 |
| $T_{Pol}$ | × | × | × | × | × |
| $T_{FE.Gen}$ | × | × | 1 | 1 | × |
| $T_{FE.Rec}$ | 1 | 1 | × | × | × |
| $T_{PUF}$ | 1 | 2 | 2 | 2 | × |
| $T_{OCECPUF}$ | × | × | × | × | 2 |
| Total | 4.79 | 5.07 | 2.95 | 2.79 | 1.09 |

TABLE VII
FEATURE-BASED CHARACTERIZATION OF OUR PROPOSED SCHEME IN COMPARISON WITH THE OTHER PUF-BASED METHODS

| Scheme | $F_1$ | $F_2$ | $F_3$ | $F_4$ | $F_5$ | $F_6$ | $F_7$ | $F_8$ | $F_9$ | $F_{10}$ | $F_{11}$ |
|---|---|---|---|---|---|---|---|---|---|---|---|
| [6] | ✓ | ✓ | ✓ | ✗ | ✓ | ✗ | ✗ | ✓ | NA | ✓ | ✗ |
| [7] | ✓ | ✓ | ✓ | ✗ | ✗ | ✗ | ✗ | ✗ | NA | ✓ | ✗ |
| [8] | ✓ | ✓ | ✓ | ✗ | ✓ | ✗ | ✗ | ✓ | NA | ✓ | ✗ |
| [9] | ✓ | ✓ | ✓ | ✗ | ✗ | ✗ | ✗ | ✓ | NA | ✗ | ✗ |
| [10] | ✓ | ✓ | ✓ | ✗ | ✓ | ✗ | ✗ | ✓ | NA | ✓ | ✗ |
| [11] | ✓ | ✓ | ✓ | ✗ | ✓ | ✗ | ✗ | ✓ | NA | ✗ | ✗ |
| [12] | ✓ | ✓ | ✓ | ✗ | ✓ | ✗ | ✗ | ✓ | NA | ✓ | ✗ |
| [13] | ✓ | ✓ | ✓ | ✗ | ✓ | ✗ | ✗ | ✓ | NA | ✓ | ✗ |
| [14] | ✓ | ✓ | ✓ | ✗ | ✓ | ✗ | ✗ | ✓ | NA | ✓ | ✗ |
| [15] | ✓ | ✓ | ✓ | ✗ | ✓ | ✗ | ✗ | ✓ | NA | ✓ | ✗ |
| [16] | ✓ | ✓ | ✓ | ✗ | ✓ | ✗ | ✗ | ✓ | NA | ✓ | ✗ |
| [17] | ✓ | ✓ | ✓ | ✗ | ✗ | ✗ | ✗ | ✓ | NA | ✗ | ✗ |
| [18] | ✓ | ✓ | ✓ | ✗ | ✓ | ✗ | ✗ | ✗ | NA | ✗ | ✗ |
| [19] | ✓ | ✓ | ✓ | ✗ | ✓ | ✗ | ✗ | ✓ | NA | ✗ | ✗ |
| [20] | ✓ | ✓ | ✓ | ✗ | ✓ | ✗ | ✗ | ✓ | NA | ✓ | ✗ |
| [21] | ✓ | ✓ | ✓ | ✗ | ✓ | ✗ | ✗ | ✗ | NA | ✗ | ✗ |
| [22] | ✓ | ✓ | ✓ | ✗ | ✓ | ✗ | ✗ | ✗ | NA | ✗ | ✗ |
| [23] | ✓ | ✗ | ✗ | ✓ | ✗ | ✗ | ✗ | ✗ | NA | ✗ | ✓ |
| [24] | ✓ | ✓ | ✓ | ✓ | ✗ | ✗ | ✗ | ✗ | NA | ✗ | ✓ |
| [25] | ✓ | ✓ | ✓ | ✓ | ✓ | ✓ | ✗ | ✓ | ✗ | ✓ | ✗ |
| [26] | ✓ | ✓ | ✓ | ✓ | ✓ | ✓ | ✗ | ✗ | ✗ | ✗ | ✓ |
| [27] | NA | NA | ✓ | ✗ | ✓ | ✓ | ✗ | ✓ | ✓ | ✗ | ✗ |
| [28] | NA | NA | ✓ | ✗ | ✓ | ✓ | ✗ | ✓ | ✓ | ✗ | ✗ |
| [29] | NA | NA | ✓ | ✗ | ✓ | ✓ | ✗ | ✓ | ✓ | ✗ | ✗ |
| [30] | ✓ | ✓ | ✓ | ✗ | ✓ | ✓ | ✓ | ✗ | ✓ | ✗ | ✗ |
| Ours | ✓ | ✗ | ✓ | ✓ | ✓ | ✓ | ✓ | ✓ | ✓ | ✓ | ✓ |

## VI. COMPARATIVE PERFORMANCE EVALUATION

The communication and computational overhead evaluation of our scheme is presented in this section and is compared with PUF-based protocols in [27]-[30]. In addition, we provide a feature-based comparison between our scheme and the related works in the literature presented in [6]-[30].

### A. Communication Overhead

The overall communication overhead of our protocol includes the maximum size of the transmitted packets, which are communicated from $SM$ to $NG$ and vice versa is computed as : $|C_j^i, V_{NG}^i, E_{NG}^i| + |F_j^i, E_j^i, V_j^i|$.

Thus, the communication cost of the authentication and data transmission phases is (3×256) = 96B and (3×256) = 96B, respectively. Therefore, the total communication overhead of our scheme is 192B. Table IV shows the comparison between our scheme and the proposed schemes in [27]-[30]. According to this table, although our proposed protocol provides two-way communication, it has a moderate communication overhead.

### B. Computational Cost

Our approach involves treating SMs as electronic devices with limited resources, in contrast to NG which is a server with significant computational power. Consequently, we only consider the computational cost on the SM side. However, our protocol also keeps the computational demands low on the NG side. To perform the cryptographic operations on SMs, we utilize the benefits of ArduinoLibs in our experiments [51], a cryptographic library on an ARM Cortex-M3 microcontroller board named AT91SAM3X8E, 512 kB flash memory, 96 kB SRAM, and a clock speed of 84 MHz, which is quite like a real-life smart meter [52]. To measure the computation cost of a PUF operation, we first implemented a 128-bit arbiter PUF with 32 MUX stages on that board. In addition, we implemented our OCECPUF, and its computational cost was almost 2.2 times greater than a simple arbiter PUF. Furthermore, for fuzzy extractors' run-time calculation, the BCH algorithm is used in the code-offset mechanism [53]. The various cryptographic primitives' execution time is presented in Table V.

Table VI shows the number of cryptographic operators used by each scheme for one protocol runtime, which $T_h, T_{Enc}$,

$T_{Dec}, T_{RNG}, T_{Pol}, T_{FE.Gen}, T_{FE.Rec}, T_{PUF}$, and $T_{OCECPUF}$ represent the execution time of one-way hash function, AES encryption, AES decryption, PRNG, polynomial generation, BCH-based sketching, BCH-based recovering, 128-bit Arbiter PUF, and our proposed SEC PUF, respectively. In our proposed scheme, $SM_j$ needs to use 8 one-way hash functions, 1 pseudo-random number generation, and 2 proposed PUF operations for each protocol execution. Therefore, the total computational cost of our protocol is (8×0.0392)+(1×0.0823)+(2×0.353)=1.09ms. The results in Table VI show that the proposed scheme in this paper dramatically outperforms the PUF-based schemes in terms of computational overhead (about 2.56 times faster than the best among the previous PUF-based schemes).

Table VII represents a feature-based comparison between the proposed scheme in this paper and the ones presented in [6]-[30] where $F_1$ to $F_{11}$ represents data confidentiality and integrity, PUF modeling attack resistance, replay attack resistance, DoS attack resistance, impersonation attack resistance, physical security, two-way communication, SM privacy and anonymity, proper error correction scheme for PUF, forward secrecy, and lightweight design, respectively. Furthermore, ✓ means a feature is provided, ✗ means a feature is not provided, and NA means not applicable. According to this table, our proposed scheme could outperform the state-of-the-art in most of the features, except providing security for used PUF against modeling attacks, since it has been proven that APUFs are vulnerable to modeling attacks. Although collecting a large set of useful challenge-response pairs is impractical for an adversary in this protocol, we aim at proposing a modeling-attack-resistant strong PUF to solve this problem in our future works.

## VII. CONCLUSION

This paper has used the advantages of an on-chip-error-correcting PUF to provide an efficient authentication protocol for secure two-way communication in the smart grid. We showed that the number of errors in OCECPUF responses plunges to zero by adding four inverters in the delay line and a simple error detection circuit. Then, by using OCECPUF in our proposed authentication protocol, we removed the need of using fuzzy extractors, which led to dramatically decrease in computational overhead. We proved that the proposed protocol stands secure in the presence of a CK-adversary and considerably outperforms the state-of-the-art in providing feature-based assessments such as physical security, forward secrecy, mutual authentication, privacy, and anonymity.